1

# Investigation into the Spread of Misinformation about UK Prime Ministers on Twitter

Junade Ali (University of Cambridge, UK)

*Abstract*—**Misinformation presents threats to societal mental well-being, public health initiatives, as well as satisfaction in democracy. Those who spread misinformation can leverage cognitive biases to make others more likely to believe and share their misinformation unquestioningly. For example, by sharing misinformation whilst claiming to be someone from a highly respectable profession, a propagandist may seek to increase the effectiveness of their campaign using authority bias. Using retweet data from the spread of misinformation about two former UK Prime Ministers (Boris Johnson and Theresa May), we find that 3.1% of those who retweeted such misinformation claimed to be teachers or lecturers (20.7% of those who claimed to have a profession in their Twitter bio field in our sample), despite such professions representing under 1.15% of the UK population. Whilst polling data shows teachers and healthcare workers are amongst the most trusted professions in society, these were amongst the most popular professions that those in our sample claimed to have.**

## I. Introduction

The presence of misinformation on social media presents several concerns to society. These effects include harming individuals' mental health [1], undermining public health initiatives [2] and impacting satisfaction with democracy [3].

Whilst it is difficult for systems to automatically differentiate between bots and real users [9], users are dependent on their own cognitive abilities to differentiate between misinformation and the truth. Several cognitive biases can impact this process and make people more likely to share misinformation by presenting it to them in a certain way. Appeals to expert opinion are one such cognitive bias [10, 11]. If the messenger of a given piece of information holds the social status of a respected expert in the subject, people are more likely to accept such information unquestioningly.

This paper seeks to understand how those who are sharing misinformation seek to present themselves to others, specifically through the professions that they claim to hold (whether purported or actual). We aim to do this by collecting Twitter account information on those who have retweeted misinformation about former UK Prime Ministers whilst they held office and systematically reviewing what professions they claim to hold in their Twitter bios.

## II. Related Work

Recent research has highlighted the potential harm that misinformation on social media poses to society. A large-scale observational study [1] of Twitter posts made during the COVID-19 pandemic found that "users who shared COVID-19 misinformation experienced approximately two times additional increase in anxiety when compared to similar users who did not share misinformation". Another study [2] used surveys to understand the impact of COVID-19 misinformation on vaccine uptake, finding "a negative relationship between misinformation and vaccination uptake rates".

Research has also shown that misinformation has an effect on satisfaction with democracy, [3] conducted an online survey in the United States which found that greater attention to political news increased the presumed influence of misinformation on others, as opposed to oneself (especially among Democrats and Independents).

[4] studied the behaviour of Twitter users before and after exposure to misinformation and found that exposure to misinformation increased tweeting (posting) frequency amongst the target group, compared with a baseline set of users. However, unfortunately, this data does not provide us with an insight into the demographic data of those sharing misinformation. Additionally, this research did not consider the role of retweets in their analysis.

[5] brings us closer to understanding why people share misinformation on Twitter. Using representative opinion polling (provided by YouGov), the authors were able to poll a representative sample of Twitter users in the United States and ask them some political questions. These political questions were then correlated with their Twitter social media activity. Using this data, the authors conclude that: "individuals who report hating their political opponents are the most likely to share political fake news and selectively share content that is useful for derogating these opponents".

Wider psychological research has sought to understand the psychological properties of those with radical beliefs. Notably, [6] has found that "individuals holding radical beliefs (as measured by questionnaires about political attitudes) display a specific impairment in metacognitive sensitivity about low-level perceptual discrimination judgments." Study participants with radical political views were less able to critically evaluate the correctness of their answers to a non-political task and, when presented with post-decision evidence against their original answer, showed reduced ability to update their confidence on the correctness of that answer.

Additionally, there has been research exploring the impact of social media on the 2016 referendum in the United Kingdom on European Union membership (resulting in what is commonly known as *Brexit*) [7] and during subsequent national General Elections [8]. Additionally, [9] has explored anti-Brexit groups within the context of bot detection. The authors collected data from accounts using the *#FBPE* (Follow Back Pro European) Twitter hashtag. The research found that it was extremely difficult to correctly distinguish between bot accounts and real accounts, with the onus being on the end-user to be able to distinguish between the two when consuming information, stating that "it would appear that for the time being at least, the onus is on the individual user to explore suspected Twitter accounts and report them if they are seen to be acting against the rules of Twitter."

Finally, behavioural psychology has long shown that

---

J. Ali is with the University of Cambridge, United Kingdom (e-mail: mjsa2@ cam.ac.uk).

appeals to expert opinion can be used to influence what people believe without critical thought [10, 11], however, unfortunately, there has been very limited research exploring either the role this cognitive bias plays in either the spread of misinformation on social media, or the exploitation of this cognitive bias by those seeking to spread misinformation.

## III. Methodology

Our goal is to gain a deeper understanding of the retweeters of misinformation about former Prime Ministers of the United Kingdom. We do this by evaluating the Twitter profiles of 1873 retweet events of tweets that have been identified as known misinformation by independent fact-checkers.

The 1873 retweet events are associated with two highly-shared tweets containing misinformation. The first tweet [12] was made on the 19th of April 2019, less than 8 months before the 2019 UK General Election. The tweet criticised the then Prime Minister, Theresa May (of the right-wing Conservative party), whilst praising the opposition leader, Jeremy Corbyn (of the left-wing Labour party). The tweet was shared with an image of Theresa May with a fake quote stating: "Curbing the promotion of lesbianism in Merton's schools starts with girls having male role models in their lives". The independent fact-checking organisation Full Fact found [13] that: “There is no evidence she ever said this.”

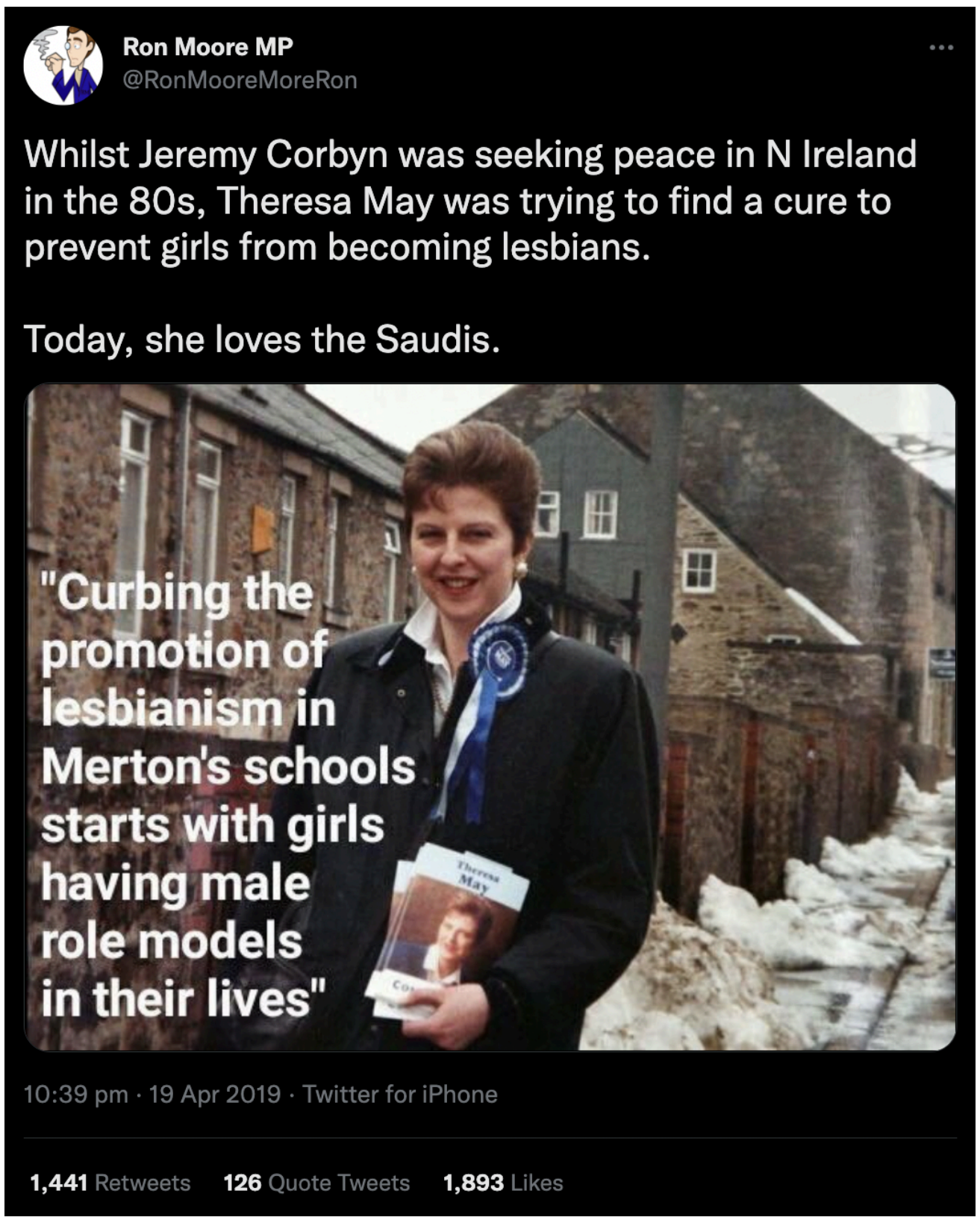

Fig. 1. Tweet [12] spreading misinformation about Prime Minister Theresa May, debunked by independent fact-checkers [13].

The second tweet [14] was sent on the 27th of March 2021 when Boris Johnson (of the right-wing Conservative party) was Prime Minister of the United Kingdom. The tweet contains an image purporting to quote Boris Johnson expressing a desire to be born in the middle ages so that he could behead “smelly peasants” and then anally rape a stable boy. The independent fact-checkers, Full Fact, again found no evidence of the former Prime Minister expressing such a desire [15].

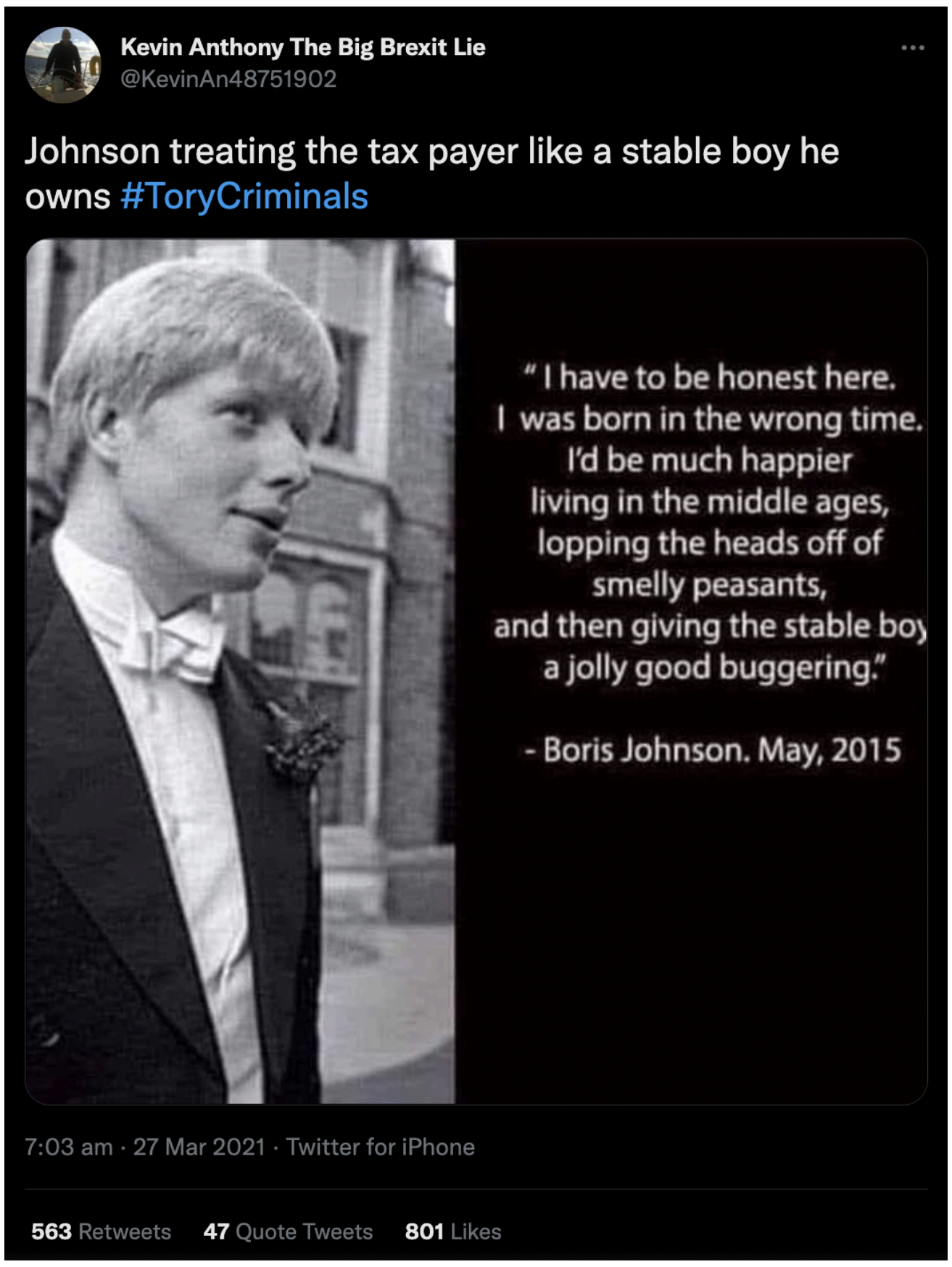

Fig. 2. Tweet [14] containing a fictitious quote about Prime Minister Boris Johnson, debunked by independent fact-checkers [15].

Using a Python environment in a Jupyter notebook, we used the Twitter API to gather the profile information of those who retweeted these two tweets. This data collection was performed on the 4th of September 2022.

We then enriched this information by collecting information about the self-reported professions of the users who shared these tweets. Using the data from user bios (sometimes known as their account description or biography), we then added an additional column of data containing the profession that the Twitter user claimed to hold.

We used a simple taxonomy which categorised the workers into the following professions:

- Creative (including artists, writers, graphic designers, poets, etc)
- Teacher/Lecturer
- Healthcare Worker (including doctors, nurses, clinical psychologists, etc)
- Engineer / Technician (including software engineers and developers)
- Businessperson
- Lawyer (including barristers and solicitors)
- General Worker (including manual labourers)
- Student
- Driver / Transportation Worker
- Salesperson / Marketer

- Soldier
- Union Official
- Journalist / Presenter
- Firefighter
- Anthropologist
- Veterinarian
- Librarian
- Accountant
- Translator
- Fisherman
- Psychic Medium
- Architect

Where a user was retired but mentioned their profession prior to retirement, we would categorise them according to their profession (many of those in the dataset were elderly and there were 59 instances of the phrase “retire” in the dataset). Students were added as a distinct category of profession.

Where a user had multiple potential professions in their bio, the first was selected (unless it was clear that another was their actual full-time job), for example; one user’s bio contained the phrase “Grief counsellor, Psychic”, so they were categorised as a healthcare worker on the basis of being a grief counsellor.

With the data enriched, in order to investigate this dataset, we firstly generated word clouds of common words in people’s profile names, bios and locations. We additionally extracted emojis from user bios and curated a list of the most frequently used ones. We further generated analysis from the data on the professions of the Twitter users sharing this misinformation.

## IV. Results & Analysis

The author of this paper subjectively found categorising the retweeters into their various professions more emotionally demanding than expected. Many of the Twitter profile bios of these users contained heartbreaking stories; for example, one user’s bio stated: *“2 failed marriages and I'm devastated 😭😭. Heartbroken 💔 over the most recent. I just want loyal, trustworthy, fun friends please.”*

Another profile bio from someone claiming to be a “grandfather” claimed they were suffering from “incurable lung cancer”. Another profile claims they were bought to creating a Twitter profile whilst suffering from cancer: *"cancer eleven and a half years ago brought me to twitter. i,ve refound my voice but the three (i,s) isolationism,ineptitude and idiocy keep me here ."*

The impact of the fear of death on people’s beliefs and actions is studied in Terror Management Theory [16]. Future research into the role (if any) that death plays in drawing people into sharing misinformation on social media could help identify the causes for this.

Additionally, many of the profiles would largely contain information claiming victimhood, for example: *“Autistic lone Mum of 2, with MS.”* ... *“#MSer #Ace #NonBinary ⚧ *She/her *They/them ✋#GTTO #Socialist"*. Studying the psychological causes behind why such Twitter profiles claim victimhood may influence ideas around the psychology of victimhood culture [17].

Also of interest is how a number of the Twitter profiles were from people who are English but choose to identify with other countries. For example, one user described themselves as a "New Scot." Another user praised their support of Scotland's secession from the United Kingdom, and supporting the pro-independence SNP (Scottish National Party): *“❤️🏴󠁧󠁢󠁳󠁣󠁴󠁿 My home for 16 years Can't wait 4 Independence from corrupt abusive WM control.English by birth SCOTTISH by choice No SNP Haters tolerated”*. (N.B. When the author says “*WM control*”, they are seemingly referring to Westminster, where the British Parliament is based in the Palace of Westminster and where many British Government institutions are located in Whitehall.)

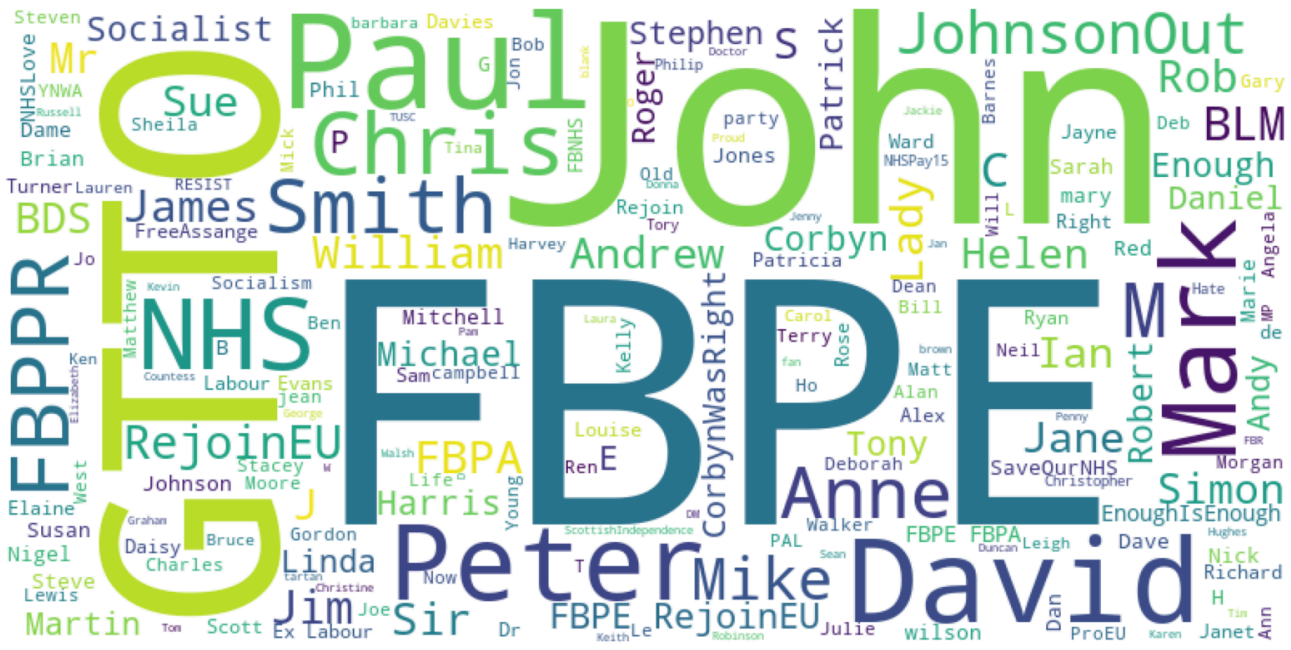

Fig. 3. Word cloud visualising the most common words appearing in the name of those who retweeted the misinformation.

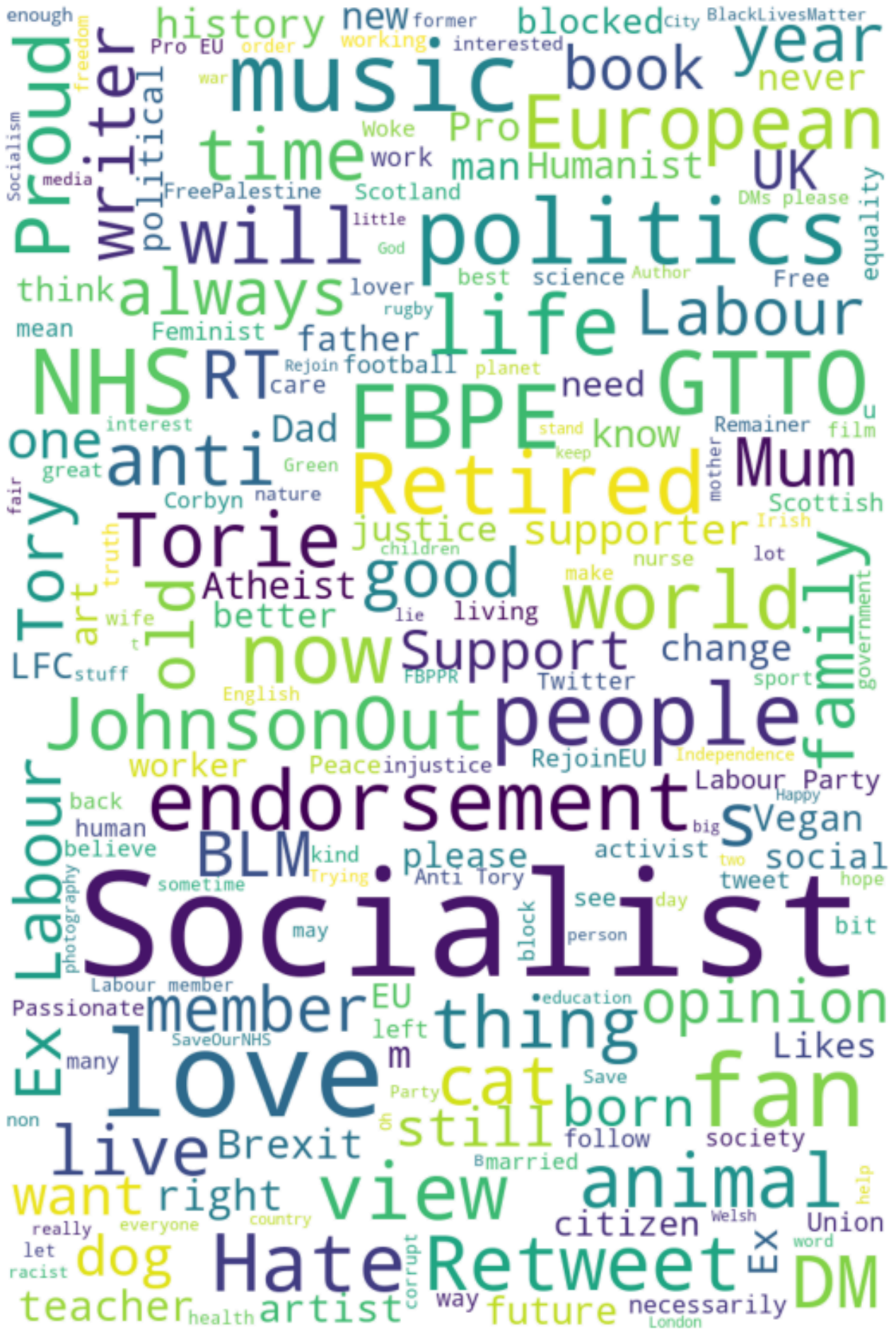

Fig. 4. Word cloud visualises the most common words appearing in the bios of those who retweeted the misinformation.

Fig. 3 shows a word cloud of the most common words appearing in the user’s profile name field, set by them. As can be seen, many users would use their name field to identify with the FBPE (Follow Back Pro European) movement [9]. Others would include hashtags like "GTTO" (Get The Tories Out), referring to the UK's Conservative party.

Fig. 4 visualises the common words listed in the Twitter bios of these users. Like the name field, many users describe themselves as "socialist" and use terms like "NHS" as they seek to praise the UK's National Health Service.

In Fig. 5 we see a word cloud of the locations set in the Twitter profiles of those whose retweets we are examining. As we can see, the users largely report being in the United Kingdom, though with some from other countries (such as Ireland and Australia). Some of these locations were set in fictional places, for example, some users used the term "Plague Island", a derogatory term for the island of Great Britain during the COVID-19 pandemic used by those opposed to the Government.

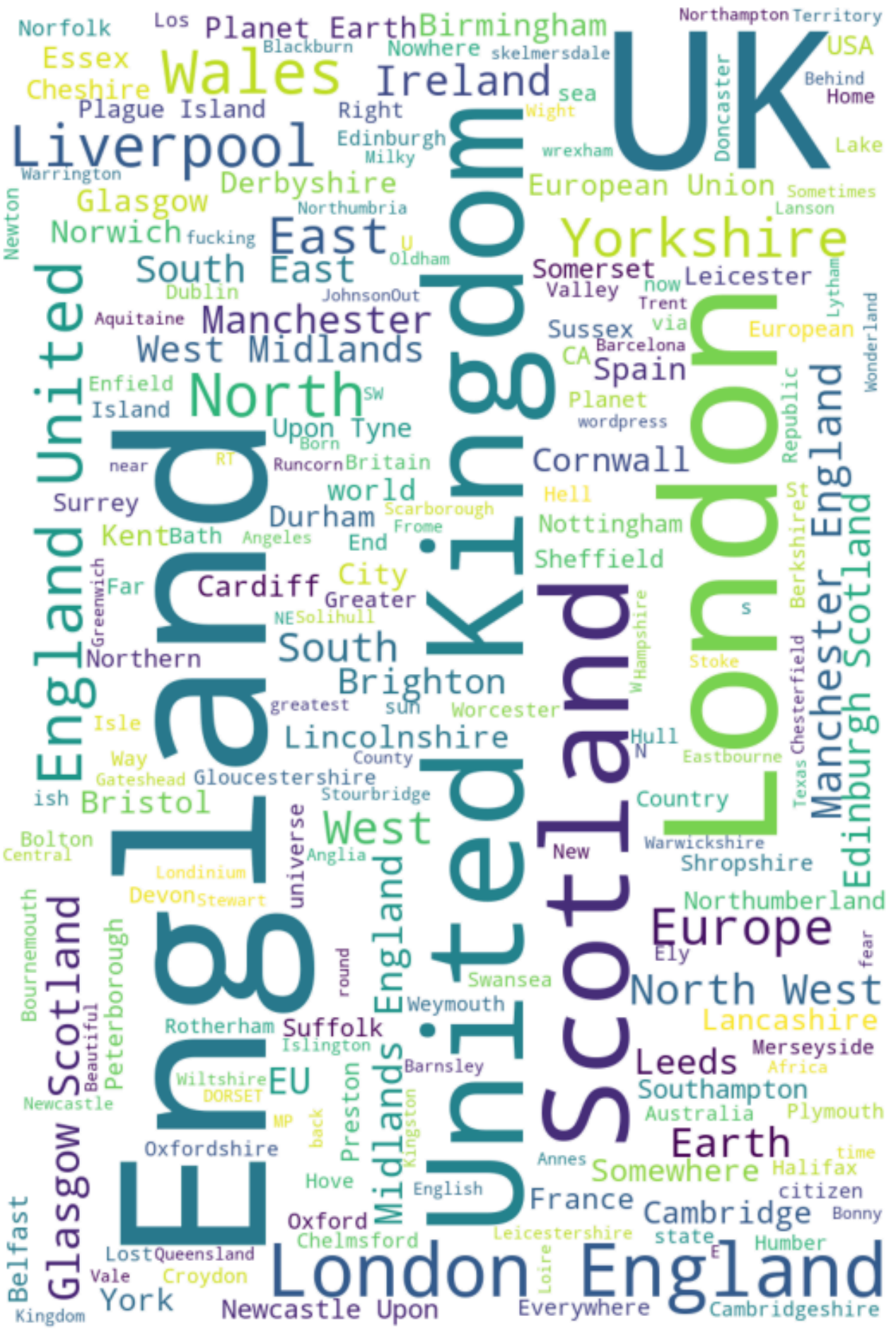

Fig. 5. Word cloud of locations set in the Twitter profiles of the retweeters.

The impact of COVID-19 can also be seen as we move to study the emojis used in the Twitter bios of these users, the top 20 of which can be seen in Table 1 alongside their respective occurrences. Note how the mask-wearing emoji appears as the 11th most common emoji in the table, but the syringe emoji (typically associated with being vaccinated against COVID-19) does not appear in the table. Indeed, the dataset only contained 3 instances of the syringe emoji being used, 1 instance by an anti-vaxer (someone opposed to vaccinations). One potential explanation for this is that the UK Government ran one of the most successful COVID-19 vaccine programs in the world (see [18] for further information on this) and those opposed to the UK Government do not want to draw attention to this.

The impact of the pandemic can also be seen in the fact that a blue heart is the most commonly used emoji in the dataset, used to symbolise support for the British National Health Service.

The second most commonly used emoji, and the most commonly used flag, was the European flag (adopted by the European Union). This was followed by the Ukrainian flag (symbolising support during the 2022 Ukraine War), followed by other flags including those of Scotland, Palestine, the UK, the LGBT rainbow flag, Wales and the Irish flag. An emoji of the world as a globe also appeared.

Other seemingly innocuous emojis have also taken on political meaning; for example, a rose symbolising the Labour party or socialism and a spider symbolising the spider broach worn by the justice of the UK Supreme Court who delivered a ruling against the Government during the 2019 British constitutional crisis over prorogation.

TABLE I: Top 20 emojis used in the Twitter profile bios of the retweeters

| Emoji | Frequency |
|---|---|
| 💙 | 203 |
| 🇪🇺 | 98 |
| 🇺🇦 | 69 |
| 🏴󠁧󠁢󠁳󠁣󠁴󠁿 | 45 |
| 🌹 | 39 |
| 🇵🇸 | 38 |
| 💚 | 33 |
| 🌻 | 29 |
| 🇬🇧 | 28 |
| 🏳️‍🌈 | 20 |
| 😷 | 18 |
| 🐟 | 16 |
| 🏴󠁧󠁢󠁷󠁬󠁳󠁿 | 16 |
| 🇮🇪 | 15 |
| 🕷️ | 15 |
| 🌍 | 15 |
| ❤️ | 12 |
| 🟨 | 12 |
| 🟥 | 12 |
| ✊ | 12 |

Turning now to the data on the professions that the users claimed to have in their Twitter bios, this information is summarised in Table 2 (limited to the top 10 professions) and this data is visualised in Fig 6.

The most popular profession appears to be creatives - however, it is important to note that we were not able to distinguish between the level of professionality involved due to the limited number of information that could be collected from a Twitter bio. For example; self-published authors, part-time musicians, etc were counted equally as graphic designers or professional filmmakers.

Interestingly, of the total dataset, 3.1% of users identified themselves as either a teacher or a lecturer. 20.7% of all those whose Twitter bio declared themselves as holding a profession were teachers or lecturers.

Data from the British Educational Suppliers Association lists that there are 624,520 full-time school teachers in the UK [19] and the UK Government's Higher Education Statistics Agency claims there are 146,780 full-time academic staff in UK higher education [20]. Given the population of the United Kingdom is estimated to be 67.1 million [21], this would mean that less than 1.15% of the UK population are teachers or lecturers (on a full-time basis). By contrast, we see that 3.1% of users who retweeted misinformation claimed to be teachers or lecturers.

There are a number of potential explanations for this. Firstly, it might be the case that whilst these users purport to be teachers and lecturers, they may simply be doing so in order to develop social proof to gain greater credibility when sharing misinformation [10, 11]. Another explanation is that if academics are more likely to oppose the political parties that UK Prime Ministers represent, a small minority sharing misinformation may be sufficient to overrepresent the profession in the sharing of misinformation.

Although healthcare workers are the third most common profession that the retweeted claimed to have, it is worth noting that the United Kingdom's National Health Service employs approximately 1.4 million people with a further 1.6 million people working in social care [22]. According to the population estimate provided in [21], this amounts to 4.47% of the UK population working in healthcare.

It is important to stress that these comparisons are crude, given we don't have a concrete baseline to compare against. Future research might wish to use representative opinion polling to identify which professions are overrepresented or underrepresented in sharing such forms of misinformation. This could also potentially be weighted by the political orientation of the misinformation being shared. Nevertheless, our work here does show that those sharing misinformation will often self-identify themselves as having a respectable profession, likely improving the spread of their misinformation by developing social proof.

The opinion polling firm, IPSOS Mori produces a Veracity Index which measures the public trust in various professions annually, the results of which show that teachers are the 4th most trusted profession measured (86% trust) behind doctors (91% trust), with nurses in 1st place (94% trust) [23].

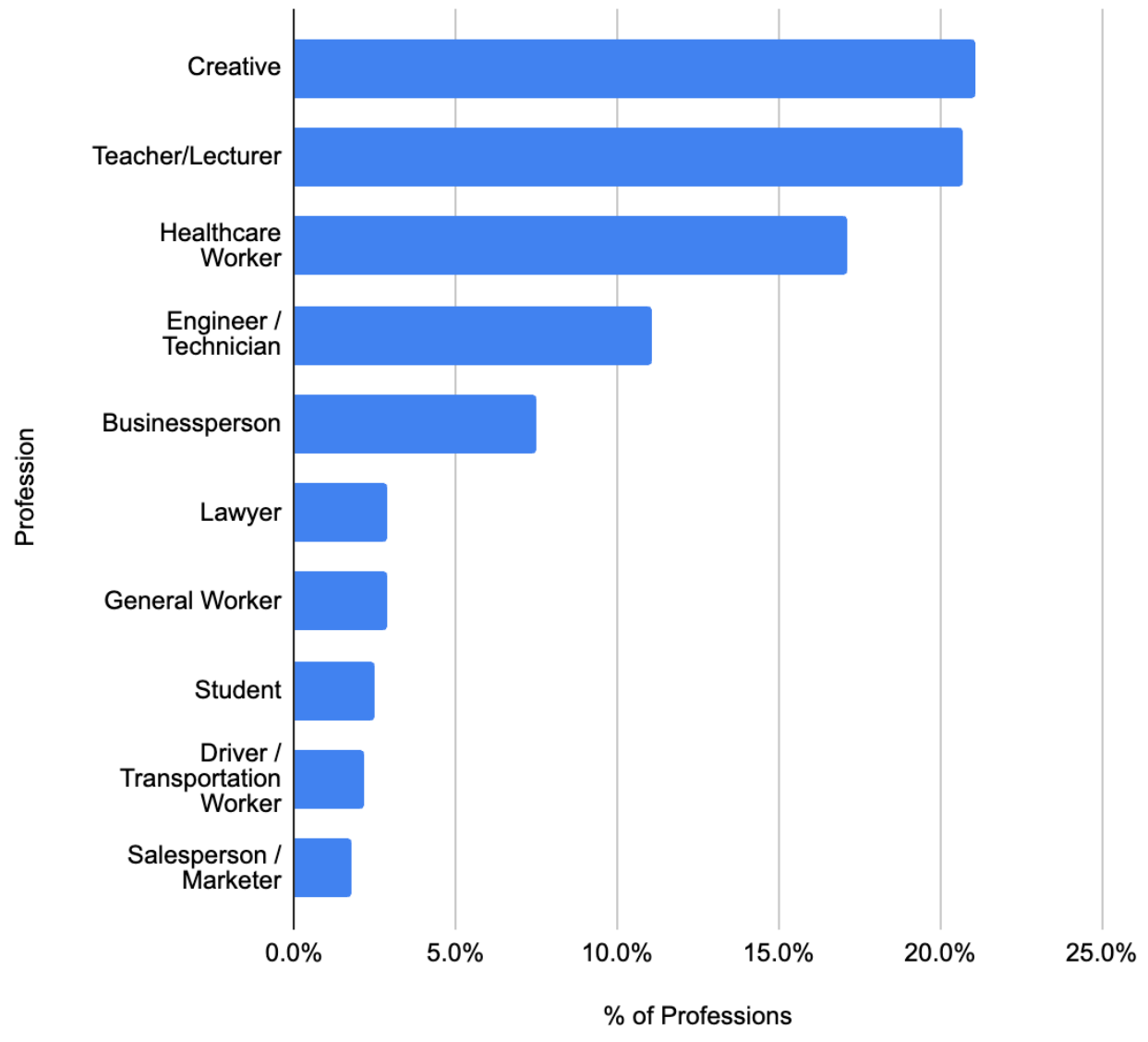

Fig. 6. Visualisation of the top 10 professions that the retweeters claimed to have in their Twitter bios.

TABLE 2: Top 10 professions that the retweeters claimed to have

| Profession | % of Professions | % of All |
|---|---|---|
| Creative | 21.1% | 3.2% |
| Teacher / Lecturer | 20.7% | 3.1% |
| Healthcare Worker | 17.1% | 2.6% |
| Engineer / Technician | 11.1% | 1.7% |
| Businessperson | 7.5% | 1.1% |
| Lawyer | 2.9% | 0.4% |
| General Worker | 2.9% | 0.4% |
| STUDENT | 2.5% | 0.4% |
| DRIVER / TRANSPORTATION WORKER | 2.1% | 0.3% |
| SALESPERSON / MARKETER | 1.8% | 0.3% |

## V. CONCLUSION

In this paper we have collected Twitter profile data about users who have retweeted known misinformation about two former UK Prime Ministers (Theresa May and Boris Johnson). We have then visualised this data using word clouds to present the most common words appearing in various profile fields. We have also identified the most common emojis used in their profile bios. Finally, we categorised these users by the professions they claim to have.

Our research has found that whilst those who spread misinformation about these two Prime Ministers typically identify as victims or as part of groups they perceive to be oppressed, a significant number also claim to have a number of respectable professions. Whilst creatives (writers, musicians, film directors, artists, etc) were the most common professions, these were followed by teachers and lecturers alongside healthcare professionals. The opinion polling firm, IPSOS Mori, has found that the public place high levels of trust in members of these professions [23].

In order to maintain both public trust in these professions, whilst also limiting the spread of misinformation it is important to develop a popular understanding of the authority bias and develop abilities in the population to critically assess content. Additionally, social media companies may wish to take measures to prevent users from purporting to be members of regulated professions when they are not, and further, professional bodies and regulators may seek to educate and implement disciplinary measures against members of a profession who spread misinformation.

Further research may wish to seek to understand empirically the proportion of each profession that is likely to spread misinformation and the different political orientation of misinformation spread by each group of professionals.

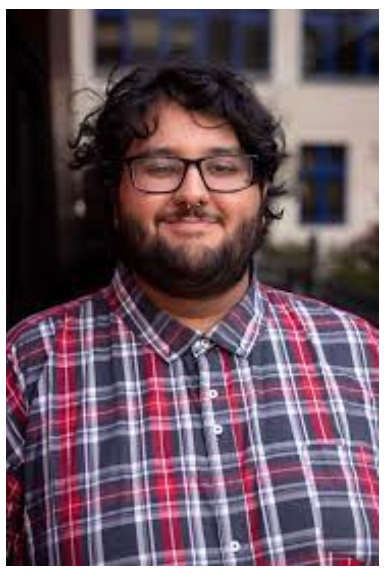

**Junade Ali** is a British computer scientist with a focus on cybersecurity, software development, cryptography and data science, and is currently studying cognitive psychology at the University of Cambridge. He is currently researching how psychology impacts political beliefs and the spread of misinformation online.